\documentstyle[12pt]{article}
\newcommand{\eq}{\begin{equation}}
\newcommand{\en}{\end{equation}}
\newcommand{\eqn}{\begin{eqnarray}}
\newcommand{\enn}{\end{eqnarray}}
\newcommand{\CR}{\nonumber \\}
\newcommand{\E}{{\rm e}}
\newcommand{\I}{{\rm i}}
\newcommand{\pa}{\partial}

\newcommand{\bvp}{\bar{\varphi}}
\newcommand{\mat}[2]{\left(\begin{array}{#1}#2\end{array}\right)}
\newcommand{\A}{\alpha}
\newcommand{\B}{\beta}
\newcommand{\D}{\delta}
\newcommand{\DE}{\Delta}
\newcommand{\G}{\gamma}
\newcommand{\ep}{\epsilon}
\newcommand{\lm}{\lambda}
\newcommand{\vp}{\varphi}
\newcommand{\cL}{{\cal L}}
\newcommand{\bg}{{\bf g}}
\newcommand{\bn}{{\bf n}}
\newcommand{\hbg}{\hat{\bf g}}
\newcommand{\hbn}{\hat{\bf n}}
\newcommand{\dpsi}{\psi^{\dagger}}
\newcommand{\TA}{\tilde{\A}}
\newcommand{\tT}{\tilde{T}}
\newcommand{\TG}{\tilde{G}}
\newcommand{\tB}{\tilde{\beta}}
\newcommand{\tG}{\tilde{\gamma}}
\newcommand{\tb}{\tilde{b}}
\newcommand{\tc}{\tilde{c}}
\newcommand{\tJ}{\tilde{J}}
\newcommand{\tj}{\tilde{j}}

\newcommand{\tH}{\tilde{H}}

\newcommand{\QB}{Q_{BRST}}
\newcommand{\cA}{{\cal A}}
\newcommand{\dyna}{\mbox{
                   \setlength{\unitlength}{1.5pt}
                   \begin{picture}(27,3)
                   \put(0,3){\circle{6}}
                   \put(0,3){\makebox(0,0){$\times$}}
                   \put(4,3){\line(1,0){15}}
                   \put(23,3){\circle{6}}
                   \put(23,3){\makebox(0,0){$\times$}}
                   \end{picture}}}
\newcommand{\dynb}{\mbox{
                   \setlength{\unitlength}{1.5pt}
                   \begin{picture}(27,3)
                   \put(0,3){\circle{6}}
                   \put(4,3){\line(1,0){15}}
                   \put(23,3){\circle{6}}
                   \put(23,3){\makebox(0,0){$\times$}}
                   \end{picture}}}

\begin{document}
\renewcommand{\thefootnote}{\fnsymbol{footnote}}
\begin{titlepage}
\null
\begin{flushright}
UTHEP-264 \\
October, 1993
\end{flushright}
\vspace{2.5cm}
\begin{center}
{\Large \bf
Hamiltonian Reduction \\
and Topological Conformal Algebra \\
in $c\leq 1$ Non-critical Strings
\par}
\lineskip .75em
\vskip 3em
\normalsize
{\large Katsushi Ito}
\vskip 1.5em
{\it Institute of Physics, University of Tsukuba, Ibaraki 305, Japan}
\vskip 1.5em
and
\vskip 1.5em
{\large Hiroaki Kanno}
\vskip 1.5em
{\it DAMTP, University of Cambridge, Cambridge, CB3 9EW, U.K.}
\vskip .5em
and
\vskip .5em
{\it Department of Mathematics, Hiroshima University,
Higashi-Hiroshima 724, Japan }\footnote{Present Address}
\vskip 1.5em
{\bf Abstract}
\end{center} \par
We study the hamiltonian reduction of affine Lie superalgebra
$sl(2|1)^{(1)}$.
Based on a scalar Lax operator formalism, we derive the free
field realization of the classical topological topological algebra
which appears in the $c\leq1$ non-critical strings.
In the quantum case, we analyze the BRST cohomology to get the
quantum free field expression of the algebra.

\end{titlepage}
\renewcommand{\thefootnote}{\arabic{footnote}}
\setcounter{footnote}{0}
\baselineskip=0.7cm
Two-dimensional topological gravity coupled to topological matter
system \cite{Wi} has been
known to reproduce the results of the matrix model approach to
two-dimensional gravity.
There is more conventional approach of considering
Liouville gravity coupled to BPZ minimal model
($c\leq1$ non-critical string) \cite{DDK}.
Although the BRST analysis of physical states \cite{LZ}
indicates that the non-critical
string theory is equivalent to other approaches, it lacks apparent
topological field theoretical formulation.

It is recently found that the BRST algebra in the non-critical string may be
extended to an infinite dimensional symmetry
(we call it topological conformal algebra) \cite{GS}.
This observation allows us to interpret the concepts in non-critical strings
such as physical states and the ground ring in terms of the topological
conformal field theory.
In a recent paper \cite{BLNW}, Bershadsky et al. suggested
that this kind of realization of topological symmetry
may be obtained from the hamiltonian reduction of affine Lie
superalgebra $sl(2|1)^{(1)}$ with a certain choice of simple root system.
The symmetry algebra is regarded as the anti-twisted $N=2$ superconformal
algebra.
It is quite important to study this correspondence
given by the hamiltonian reduction and examine
topological properties in view of its extension
to the non-critical $W$-strings, which have rich
symmetries and gives well-defined continuum models
beyond the $c=1$ barrier.

The purpose of the present article is to study the classical and quantum
hamiltonian reduction method of $sl(2|1)^{(1)}$ relevant to the $c\leq 1$
non-critical string in detail.
After introducing the affine Lie superalgebra $sl(2|1)^{(1)}$ and
two types of simple root system, we show that
the Poisson algebra on the reduced phase space gives the
(twisted) $N=2$ superconformal algebra.
We then propose a scalar Lax operator formalism to get a free field
realization of the classical topological conformal algebra,
which is different from the standard one in the sense that
one of the super currents is realized by a single fermionic
(anti-ghost) field.
We shall study the quantum hamiltonian reduction of
$sl(2|1)^{(1)}$ based on the BRST formalism.
There are two approaches to study the BRST cohomology.
One is to use an explicit Wakimoto realization of the currents and to
study the quartet made of  bosonic and fermionic ghosts \cite{BeOo}.
The other approach first studied by Feigin and Frenkel \cite{FeFr}
uses the canonical coboundary operator of a nilpotent subalgebra
of affine Lie algebra, which does not rely on any explicit free field
realization of the currents. In the present article we will use the
latter method. In the quantum hamiltonian reduction the
topological conformal algebra is obtained as the
algebra of BRST cohomologies. After bosonizing the Cartan currents,
we get the free field realization of topological conformal
algebra in the non-critical string theory.

Let $\bg$ be a Lie superalgebra with rank $r$,
$\hbg$ its untwisted affine extension \cite{Kac}.
Denote the set of roots of $\bg$ by $\DE=\DE_{+}\cup\DE_{-}$, where
$\DE_{+}$ ($\DE_{-}$) is the set of positive (negative) roots of $\bg$.
The set $\DE$ may be also decomposed as $\DE^{0}\cup\DE^{1}$.
Here $\DE^{0}$ ($\DE^{1}$) is the set of even (odd) roots, which
correspond to the (anti-)commuting generators.
Define $\DE^{i}_{+}$ ($i=0,1$) by $\DE^{i}\cap\DE_{+}$.

The Lie superalgebra $sl(2|1)$ is represented by $3\times 3$ matrices
\eq
X=\mat{cc|c}{x_{11}& x_{12} & \xi_{13}\cr
             x_{21}& x_{22} & \xi_{23} \cr
             \hline
             \xi_{31} & \xi_{32} & x_{33}},
\label{eq:elm}
\en
satisfying the super traceless condition
${\rm str} X\equiv x_{11}+x_{22}-x_{33}=0$.
Here $x$'s and $\xi$'s are Grassmann even and odd elements, respectively.

In contrast to simple Lie algebras, a Lie superalgebra admits in general
various choices of the simple root system.
For $\bg=sl(2|1)$ there are two distinct simple root systems.
One is given by
\eq
\A_{1}=e_{1}-e_{2}, \quad \A_{2}=e_{2}-\D_{1},
\label{eq:typea}
\en
which corresponds to the Dynkin diagram \dynb .
Here $e_{i}$ $(i=1,2)$ and $\D_{1}$ are orthonormal basis with positive
and negative metric respectively: $e_{i}\cdot e_{j}=\D_{i,j}$,
$\D_{1}\cdot\D_{1}=-1$.
In the choice (\ref{eq:typea}),
the simple roots $\A_{1}$ and $\A_{2}$
satisfy  $\A_{1}^{2}=2$, $\A_{2}^{2}=0$ and $\A_{1}\cdot\A_{2}=-1$.
The positive roots of this algebra are $\A_{1}$, $\A_{2}$ and
$\A_{1}+\A_{2}$, in which $\A_{1}$ is an even root and
others are odd roots.
The other choice is to take the purely odd simple root system:
\eq
\TA_{1}=e_{1}-\D_{1}, \quad \TA_{2}=\D_{1}-e_{2},
\label{eq:typeb}
\en
which corresponds to the Dynkin diagram \dyna .
Both root systems are related each other by
\eq
\A_{1}=\TA_{1}+\TA_{2}, \quad \A_{2}=-\TA_{2}.
\en
We call the former root system as type A and the latter as type B.

The type B root system has been  used to derive
the $N=2$ superconformal $CP_{1}$ model by gauging the nilpotent
subalgebra of $sl(2|1)^{(1)}$ \cite{BeOo2}.
In this case we can use manifest $N=1$ supersymmetric formulation since
the embedding of $sl(2)$ may be extended to Lie superalgebra $osp(1|2)$
\cite{EvHo}.
On the other hand, the type A simple root system does not allow
manifest superspace formulation.
Hence we must use component formalism for the A type.
But it is the reduction based on the type A simple root system
which gives the topological conformal algebra of the
$c \leq 1$ non-critical string \cite{BLNW}.
Note that the nilpotent subalgebra $\hbn^{*}$ depends
on the choice of simple root system.

The dual space $\hbg^{*}$ of an affine Lie algebra $\hbg$ carries a
Poisson bracket structure induced by the coadjoint action of $\hbg$,
or the gauge transformation.
The space $\hbg^{*}$ is generated by
bosonic currents $J_{\A}(z)$ ($\A\in\DE^{0}$), $H^{i}(z)$
($i=1,\ldots,r$) and
fermionic currents $j_{\G}(z)$ ($\G\in\DE^{1}$).
For simplicity we use the notations
$J_{\pm 1}(z)$, $j_{\pm 2}(z)$, $j_{\pm 12}(z)$ for the currents
corresponding to the roots
$\pm\A_{1}$,$\pm\A_{2}$ and $\pm(\A_{1}+\A_{2})$.
The current algebra is expressed in the form of the operator
product expansions (OPEs):
\eqn
J_{\pm 1}(z)j_{\mp 12}(w)&=&{\mp j_{\mp 2}(w)\over z-w}+\cdots, \quad
J_{\pm 1}(z)j_{\pm 2}(w)={\pm j_{\pm 12}(w)\over z-w}+\cdots, \CR
j_{\pm 2}(z)j_{\mp 12}(w)&=& {-J_{\mp 1}(w)\over z-w}+\cdots, \CR
J_{\A}(z)J_{-\A}(w)&=&{k\over (z-w)^{2}}+{\A\cdot H(w)\over z-w}+\cdots,
\quad\mbox{for $\A\in\DE^{+}_{0}$}, \CR
j_{\A}(z)j_{-\A}(w)&=&{-k \over (z-w)^{2}}+{-\A\cdot H(w)\over z-w}
+\cdots, \quad\mbox{for $\A\in\DE^{+}_{1}$},\CR
H^{i}(z)J_{\A}(w)&=& {\A^{i}J_{\A}(w)\over z-w}+\cdots, \quad
H^{i}(z)j_{\A}(w)={\A^{i}j_{\A}(w)\over z-w}+\cdots, \CR
H^{i}(z)H^{j}(w)&=& {k\D_{i j}\over (z-w)^{2}}+\cdots.
\enn

We now study the classical hamiltonian reduction of affine
Lie superalgebra $sl(2|1)^{(1)}$.
The hamiltonian structure of the $N=2$ superconformal algebra
may be realized on the reduced phase space
$\hbg^{*}\otimes \{ \dpsi , \psi \}/\hbn^{*}$,
where $\dpsi(z)$ and $\psi(z)$ are  a pair of fermionic ghosts with weight
$\lm$ and $1-\lm$ introduced for making the second class constraints into
the first class constraints.
The ghost pair $\dpsi(z)$ and $\psi(z)$ carries the Cartan weights
$-\A_{2}$and $\A_{2}$  ($-\TA_{2}$ and $ -\TA_{1}$) for type A (B).
The weights of auxiliary fermionic fields are determined by the gradings
of the $sl(2)$ embedding into $\bg$ and depend on the choice of the root
system ($\lm=3/2$ for type A and $\lm=1/2$ for type B ).

In both cases of simple root systems, one may take the Drinfeld-Sokolov
(DS) gauge:
\eq
J_{DS}(z)=\mat{cc|c}{ {U(z)\over2} & 1 & 0 \cr
                      \tT(z)  & {U(z)\over2} & \TG^{+}(z) \cr
                       \hline
                       \TG^{-}(z)  & 0 & U(z)}.
\en
The gauge transformation $\D_{\ep}J_{DS}(z)=[\ep(z), J_{DS}]+k\pa\ep(z)$
preserving the DS-gauge, induces the Poisson algebra structure on the
reduced phase space.
Put the gauge parameter $\ep(z)$ to be
\eq
\ep(z)=\mat{cc|c}{ \ep_{11}+{x\over2}& \ep_{12} & \xi_{1} \cr
                   \ep_{21}     & -\ep_{11}+{x\over2} & \xi_{2} \cr
                       \hline
                   \eta_{1}  & \eta_{2} & x}.
\en
{}From the DS-gauge preserving condition, we get the relation
\eqn
\ep_{11}&=& -{k\over2}\pa\ep_{12}, \CR
\ep_{21}&=& \tT\ep_{12}-{k^{2}\over2}\pa^{2}\ep_{12}
            +{1\over2}(\eta_{2}\TG^{+}-\TG^{-}\xi_{1}), \CR
\eta_{1}&=& \TG^{-}\ep_{12}+{U\over2}\eta_{2}-k\pa\eta_{2}, \CR
\xi_{2}&=& {U\over2}\xi_{1}+\TG^{+}\ep_{12}+k\pa\xi_{1}.
\enn
The gauge parameter with this restriction
defines the gauge transformation on the reduced phase space.
By expressing its generator $\D$ as
\eqn
\D&=&\int {d z \over 2 \pi\I}{\rm str} \biggl\{
\mat{cc|c}{ {U(z)\over2} & 0 & 0 \cr
                      \tT(z)  & {U(z)\over2} & \TG^{+}(z) \cr
                       \hline
                       \TG^{-}(z)  & 0 & U(z)}
\mat{cc|c}{ {x\over2}& \ep_{12} & \xi_{1} \cr
            0     & {x\over2} & 0 \cr
                       \hline
            0  & \eta_{2} & x} \biggr\} \CR
&=&\int {d z \over 2 \pi\I}
 \big\{ \ep_{12}\tT-{x\over2}U+\xi_{1}\TG^{-}-
        \eta_{2}\TG^{+} \big\},
\enn
the gauge transformation is given by taking the Poisson
bracket with $\D$. Thus we find the Poisson algebra which
takes the following form in terms of the operator
product expansions:
\eqn
T_{PF}(z)T_{PF}(w)&=& {-3k\over (z-w)^{4}}
                +{2T_{PF}(w)\over (z-w)^{2}}
                +{\pa T_{PF}(w)\over z-w}+\cdots, \CR
T_{PF}(z)G^{\pm}(w)&=& {{3\over2}G^{\pm}(w)\over (z-w)^{2}}
    +{(\pa G^{\pm})(w)\pm{1\over 2k} (U G^{\pm})(w) \over z-w}+\cdots, \CR
U(z)U(w)&= & {-2k \over (z-w)^{2}}+\cdots, \CR
U(z)G^{\pm}(w)&=& {\pm G^{\pm}(w)\over z-w}+\cdots, \CR
G^{+}(z)G^{-}(w)&=&{-2k \over (z-w)^{3}}
                    +{ U(w)\over (z-w)^{2}}
                    +{T_{PF}(w)-{U^{2}(w)\over 4k}+{\pa U(w)\over 2}
                      \over z-w}+\cdots,
\enn
where $T_{PF}(w)=\tT(w)/k$, $G^{+}(w)=\TG^{+}(w)$ and $G^{-}(w)=-\TG^{-}/k$.
The energy-momentum tensor $T_{PF}$ is nothing but that of the $Z_{k}$
parafermions $SU(2)_{k}/U(1)$\cite{ZaFa}.
In fact, the total energy-momentum tensor
\eq
T^{N=2}(z)=T_{PF}(z)-{U(z)^{2}\over 4k}
\en
becomes that of $N=2$ superconformal algebra.

Next we consider a system of differential equations associated with the
Lax operator $k\pa-J_{DS}(z)$:
\eq
(k\pa -J_{DS}(z))v(z)=0, \quad
v(z)=\mat{c}{v_{1}(z)\cr v_{2}(z) \cr\hline v_{3}(z)},
\en
The top component $v_{1}(z)$ of $v(z)$ is invariant under the gauge
transformation of the unipotent subgroup which corresponds to
the nilpotent subalgebra $\hat{\bn}$.
Hence the system of equations may be written in the
form of a gauge-invariant pseudo-differential
equation of the form $\cL_{DS} v_{1}(z)=0$,
where $\cL_{DS}$ is the scalar Lax operator:
\eq
\cL_{DS} = (k\pa)^{2}-U k\pa-k T^{twisted}
          -\TG^{+}(k \pa-U)^{-1}\TG^{-},
\label{eq:ds}
\en
where
$T^{twisted}=T^{N=2}+{\pa U\over 2}$ is the twisted energy-momentum tensor.

Note that $\cL_{DS}$ is a covariant operator
under the spectral flow.
Namely for any function $\A(z)$, the scalar Lax operator
$\cL_{DS}=\cL_{DS}(U, G_{+}, G_{-}, T^{N=2})$ satisfies
\eq
\E^{\A(z)}\cL_{DS}(U, G_{+}, G_{-}, T^{N=2})\E^{-\A(z)}
=\cL_{DS}(U^{\A}, G_{+}^{\A}, G_{-}^{\A}, T^{N=2,\A}).
\en
where
\eqn
U^{\A}(z)&=& U(z)+2k\pa\A(z), \CR
G_{\pm}^{\A}(z)&=& \E^{\mp \A(z)}G_{\pm}(z), \CR
T^{N=2,\A}(z)&=& T^{N=2}(z)-k (\pa\A(z))^{2}-U(z)\pa\A(z)+k\pa^{2}\A(z).
\enn
The present formula for the spectral flow would be useful to investigate
the supersymmetry in higher spin conserved currents, which is not manifest
in the component formalism.

The free field realization is obtained by connecting the DS gauge and
the diagonal gauge\cite{DrSo}:
\eq
J_{diag}(z)=\mat{cc|c}{ P_{1}(z) & 1 & 0 \cr
                        0 & P_{2}(z) & j_{-2}(z) \cr \hline
                        0 & j_{2}(z) & P_{3}(z)},
\label{eq:diaggauge}
\en
where $P_{i}(z)=h_{i}\cdot H(z)$ ($i=1,2,3$) and
$h_{i}$ are the weights in the vector representation:
$h_{1}=\lm_{1}$, $h_{2}=\lm_{2}-\lm_{1}$, $h_{3}=\lm_{2}$.
Here $\lm_{1}\equiv -\A_{2}$ and $\lm_{2}\equiv -\A_{1}-2\A_{2}$ are
the fundamental weights satisfying $\A_{i}\cdot\lm_{j}=\D_{i,j}$.
Let us consider the system of differential equation associated with
the diagonal gauge:
\eq
(k\pa - J_{diag})v'(z)=0, \quad
v'(z)=\mat{c}{v'_{1}(z)\cr v'_{2}(z)\cr\hline v'_{3}(z)}.
\label{eq:diageq}
\en
As in the case of the DS-gauge, the above equations are shown to be
equivalent to
$\cL_{diag}v'_{1}=0$,
where
\eqn
\cL_{diag}&=&(k\pa -P_{2})(k \pa -P_{1})
 -j_{-2}(k\pa-P_{1}-P_{2})^{-1}j_{2}(k\pa-P_{1}) \CR
 &=& (k\pa)^{2}-(P_{1}+P_{2})k\pa +P_{2}P_{1}-k(\pa P_{1}) \CR
 & &  -j_{-2}j_{2}-j_{-2}(k\pa -P_{1}-P_{2})^{-1}(-k(\pa j_{2})+P_{2}j_{2}).
\label{eq:dia}
\enn
Since the top components of $v$ and $v'$ are gauge invariant,
the equations $\cL_{DS} v_{1}(z)=0$ and $\cL_{diag}v'_{1}(z)=0$ should
give the same equation.
Therefore we get the Miura transformation of the form:
\eqn
U(z)&=& P_{1}(z)+P_{2}(z), \CR
G^{+}(z)&=& j_{-2}(z), \CR
G^{-}(z)&=& \pa j_{2}(z)-{1\over k}P_{2}(z)j_{2}(z), \CR
T^{twisted}(z)
&=& -{P_{2}(z)P_{1}(z)\over k}+\pa P_{1}(z)+{1\over k}j_{-2}(z)j_{2}(z).
\label{eq:gen}
\enn

In the diagonal gauge (\ref{eq:diaggauge}), the fields
$P_{i}(z)$ ($i=1,2$) and $j_{\pm 2}(z)$ are not free fields due to the
constraints:
\eqn
j_{2}(z)j_{-2}(w)&=&{-k \over (z-w)^{2}}+{-\A_{2}\cdot H(w)\over z-w}
+\cdots, \CR
 H(z)j_{\pm 2}(w)&=&{\pm\A_{2} j_{\pm 2}(w)\over z-w}+\cdots .
\label{eq:const}
\enn
However, by introducing two free bosons
$\phi(z)=(\phi_{1}(z),\phi_{2}(z))$ and a pair of fermionic
ghosts $(\dpsi(z),\psi(z))$ with weights $({3\over2}, -{1\over2})$,
we may solve the constraints.
Define complex boson $\vp(z)=(\A_{1}+\A_{2})\cdot\phi(z)$ and
$\bvp(z)=-\A_{2}\cdot\phi(z)$.
Then a solution is given by
\eqn
j_{-2}(z)&=& \dpsi(z), \quad\quad
j_{2}(z)= k\pa \psi(z) +\I\sqrt{k} \pa \bvp(z) \psi(z), \CR
P_{1}(z)&=& \I\sqrt{k}\pa\bvp(z), \quad\quad
P_{2}(z)= -\I\sqrt{k}\pa\vp(z)+\dpsi \psi(z) .
\label{eq:free}
\enn
{}From (\ref{eq:gen}) and (\ref{eq:free}) we get
\eqn
U(z)&=& \I \sqrt{k}\pa (\bvp-\vp)(z)+\dpsi \psi(z), \CR
G^{+}(z)&=& \dpsi(z), \CR
G^{-}(z)&=& -k \pa^{2}\psi(z)+\I\sqrt{k}\pa[ \pa(\bvp-\vp) \psi](z)
 +\psi \{ -\pa\vp\pa\bvp +\I\sqrt{k}\pa^{2}\vp -\dpsi\pa \psi \}(z), \CR
T^{twisted}(z)
&=& -\pa\vp\pa\bvp(z) +\I\sqrt{k}\pa^{2}\bvp(z) -\dpsi\pa \psi(z).
\label{eq:gen1}
\enn
The fermionic ghosts $(\dpsi,\psi)$ have weights $(1,0)$ with respect to the
energy-momentum tensor $T^{twisted}$ but $(2,-1)$ for the anti-twisted
one $T^{anti-twisted}=T^{N=2}(z)-{\pa U(z)\over 2}$:
\eq
T^{anti-twisted}(z)= -\pa\vp\pa\bvp(z) +\I\sqrt{k}\pa^{2}\vp(z)
                    -2 \dpsi\pa \psi (z)
                    -(\pa \dpsi) \psi(z).
\en

Next we discuss the quantum hamiltonian reduction based on the
BRST formalism.
Let us introduce a pair of fermionic ghosts $(\tb_{1},\tc_{1})$ with
conformal weights $(0,1)$ for the
constraint for $J_{-1}(z)-1=0$ and two pairs of bosonic ghosts
$(\tB_{2},\tG_{2})$, $(\tB_{12},\tG_{12})$ with weights
$({3\over 2},-{1\over2})$, $({1\over2},{1\over2})$ for the constraints
$j_{-2}(z)-\dpsi(z)=0$ and $j_{-12}(z)=0$, respectively.
Define the BRST current $J_{BRST}(z)$ and the BRST charge $Q_{BRST}$ by
\eq
J_{BRST}(z)=\tc_{1}(J_{-1}-1)+\tG_{2}(j_{-2}-\dpsi)+\tG_{12}j_{-12}
            +\tc_{1}\tG_{2}\tB_{12},
\en
and $Q_{BRST}=\int {d z \over 2\pi \I}J_{BRST}(z)$,
satisfying the nilpotency condition $Q_{BRST}^{2}=0$.

In Feigin-Frenkel's approach \cite{FeFr}, one decomposes the BRST
current $J_{BRST}(z)$ as $J_{BRST}^{0}+J_{BRST}^{1}$, where
\eqn
J_{BRST}^{0} &=&\tc_{1}J_{-1}+\tG_{2}j_{-2}+\tG_{12}j_{-12}
             +\tc_{1}\tG_{2}\tB_{12}, \CR
J_{BRST}^{1} &=& -\tc_{1} -\tG_{2} \dpsi.
\enn
We define the associated BRST charges $\QB^{0}$ and $\QB^{1}$ by contour
integration.
$\QB^{0}$ is the canonical coboundary operator for the nilpotent subalgebra
$\hat{\bf n}_{+}$.
$\QB^{1}$ defines the resolution induced by the (first class)
constraints. Both $\QB^{0}$ and
$\QB^{1}$ act on the operator algebra $\cA_{tot}$ of the tensor product
of the current algebra and the ghost algebras;
\eq
\cA_{tot}=U\! \big[ sl(2|1)^{(1)}\big]\otimes
 Cl_{\dpsi,\psi}\otimes Cl_{\tb_{1},\tc_{1}}
          \otimes H_{\tB_{2},\tG_{2}}\otimes H_{\tB_{12},\tG_{12}},
\en
where $U\! \big[ sl(2|1)^{(1)}\big]$ is the universal enveloping algebra of
$ sl(2|1)^{(1)}$ and $Cl_{b,c}$ denotes the Clifford algebra of fermionic
ghosts $(b,c)$.
$H_{\B,\G}$ is the Heisenberg algebra of bosonic ghosts
$(\B,\G)$.
Since $\big( \QB^{0} \big)^2 = \big( \QB^{1} \big)^2
= \QB^{0} \QB^{1} + \QB^{1} \QB^{0} =0$,
the BRST cohomology may be calculated by considering
the double complex of $\QB^{0}$ and $\QB^{1}$.
Feigin and Frenkel originally computed $\QB$ cohomology by
taking $\QB^{0}$ cohomology first;
\eq
H_{\QB}(\cA_{tot})=H_{\QB^{1}}(H_{\QB^{0}}(\cA_{tot})).
\en
But, as was pointed out by de Boer and Tjin \cite{BoTj}, we can exchange
the order of taking cohomology
\eq
H_{\QB}(\cA_{tot})=H_{\QB^{0}}(H_{\QB^{1}}(\cA_{tot})),
\en
because the non-trivial $\QB^{1}$ cohomology only
appears at the ghost number zero sector (the spectral
sequence degenerates at the $E_1$ term).
\par
In practice, it is convenient to introduce
the modified currents:
\eqn
\tJ_{-1} &=&J_{-1}+\tG_{2}\tB_{12}, \quad
\tj_{-2}=j_{-2}+\tc_{1}\tB_{12},  \quad \tj_{\pm 12}=j_{\pm 12}, \CR
\tJ_{1} &=& J_{1}+\tG_{12}\tB_{2}, \quad
\tj_{2}=j_{2}+\tG_{12}\tb_{1}, \CR
\tH &=& H+\A_{1}\tc_{1}\tb_{1} + \A_{2}\tG_{2}\tB_{2}
      +(\A_{1}+\A_{2})\tG_{12}\tB_{12}.
\enn
Being BRST doublets, $(\tJ_{-1},\tb_{1})$, $(\tj_{-2},-\tB_{2})$ and
$(\tj_{-12},-\tB_{12})$ decouple from the
non-trivial cohomology.
On the reduced complex $\cA_{red}$, the non-trivial
$\QB^{0}$ action reads,
\eqn
\QB^{0}(\tJ_{1}) &=& (k+1) \partial\tc_{1} - (\tH_{1}\tc_{1}), \CR
\QB^{0}(\tj_{2}) &=& (k+1) \partial\tG_{2} - (\tH_{2}\tG_{2}), \CR
\QB^{0}(\tj_{12}) &=& (k+1) \partial\tG_{12} - (\tH_{1}+\tH_{2})
\tG_{12} -(\tj_{2}\tc_{1})_{AS} -(\tJ_{1}\tG_{2})_{S}, \CR
\QB^{0}(\tG_{12}) &=& \tc_{1}\tG_{2}.
\enn
Here we define the normal ordered product $(AB)(z)$ for two operators
$A(z)$ and $B(z)$ by $(AB)(z)=\int_{z}{dw\over 2\pi\I}{A(w)B(z)\over w-z}$.
$(AB)_{S}$ ( $(AB)_{AS}$)  denotes (anti-)symmetrized normal ordered product.
Note that the Cartan currents $H_i =  \A_i \cdot H$ have been modified
such that $\QB^{0}(\tH_{i})=0$.
Let us consider the $\QB^{1}$ cohomology first. It is
easy to see that $H_{\QB^{1}}(\cA_{red})$ is generated by
the following elements;
\eqn
U^{(0)}~&=& \dpsi\psi + (\tH_{1} + \tH_{2}), \CR
G_+^{(0)}~&=& \dpsi , \CR
G_-^{(2)}~&=& -{1\over k+1} \tj_{12} , \CR
T^{(1)}~&=& {1\over k+1} \big( \tJ_{1} +\tj_{2}\dpsi \big).
\enn
The standard argument of the spectral sequence tells us
that the generators of the total BRST cohomology are
obtained by solving the descent equation
\eq
\QB^{0} ( {\cal O}^{(n)} ) + \QB^{1} ( {\cal O}^{(n+1)} )
=0.
\en
The solvability of the descent equation is assured by the fact that
the spectral sequence degenerates at the $E_1$ term. We find
\eqn
U &=& U^{(0)}, \CR
G_+ &=& G_+^{(0)}, \CR
G_- &=& G_-^{(2)} + G_-^{(1)} + G_-^{(0)}, \CR
T &=& T^{(1)} + T^{(0)},
\enn
where
\eqn
G_-^{(1)} &=& {1\over (k+1)} \big( \tH_{1} + \tH_{2} \big)\tj_{2}
            + \psi T^{(1)} -{k\over (k+1)} \partial\tj_{2}, \CR
G_-^{(0)} &=& \psi T^{(0)} + \partial\psi \big(
               \tH_{1} + 2\tH_{2} \big) - (k+{1\over 2})
              \partial^2\psi, \CR
T^{(0)} &=& {1\over (k+1)} \bigl[ k\partial\tH_{2}
           -(\tH_{1} +\tH_{2}) \tH_{2} \bigr]
           + {1\over 2} (\partial\psi\dpsi - \dpsi\partial\psi),
\enn
The OPE of $U,~G_{\pm}$ and $T$ gives the \lq\lq twisted"
$N=2$ superconformal algebra where the twist of the stress tensor
is defined by $T=T^{N=2} + {1\over 2}\partial U$.
The central charge is $d=c/3= -(2k+1)$. The singular limit
$k \longrightarrow -1$ corresponds to $d=1$.
\par
Since the OPE respects the grading of the decomposition of each
generator, the zero grading pieces also define the algebra
isomorphic to the original algebra. Let us bosonize the modified
current as follows;
\eqn
\tH_{1} +\tH_{2} &=& \I\sqrt{k+1}\partial\vp, \CR
\tH_{2} &=& -\I\sqrt{k+1}\partial\bvp ,
\enn
and $\vp={1\over\sqrt{2}}(\phi_{1}+\I\phi_{2})$ and
$\bvp={1\over\sqrt{2}}(\phi_{1}-\I\phi_{2})$.
It is not the original current $H_i$, but the modified currents
$\tH_i$ that are bosonized. The ghost fields have non-trivial
OPE with $\phi_1$ and $\phi_2$. We get the following free
field realization of the anti-twisted $N=2$ superconformal algebra;
\eqn
U^{(0)}~&=& \dpsi\psi +\I\sqrt{2(k+1)} \partial\phi_1, \CR
G_+^{(0)} ~&=& \dpsi \CR
G_-^{(0)} ~&=& \psi T^{(0)} + \I\sqrt{2(k+1)} \partial\psi
\partial\phi_1 - (k+{1\over 2})\partial^2\psi, \CR
T^{(0)} ~&=& \partial\psi\dpsi - \partial\vp
\partial\bvp + \I{k\over\sqrt{k+1}}\partial^2\bvp .
\enn
After the twist $T^{anti-twisted}= T^{(0)} - \partial U^{(0)}$,
we recognize the free field realization of the topological
conformal algebra appearing in the non-critical string theory\cite{GS,BLNW}.
In the above equations the matter field $\phi_{M}$ and the  Liouville field
$\phi_{L}$ are identified as $\phi_{1}$ and $\phi_{2}$, respectively.
In the classical limit $k\rightarrow \infty$ these formulas reduce
to (\ref{eq:gen1}).
\par
We have shown that the BRST cohomology for the Hamiltonian
reduction of $sl(2\vert 1)^{(1)}$ by the constraints
$J_{-1}=1$ and $j_{-2}=\dpsi$ gives exactly the twisted $N=2$
superconformal algebra of the non-critical string theory.
As we remarked before, one could obtain the same result
(modulo BRST exact terms) by taking $\QB^{0}$ cohomology
first. The equivalence of the two ways of constructing
the spectral sequence allows us to get some insight
on the screening operator necessary for the free field
approach to the non-critical string theory.
The point is that we can think of $\QB^{1}$ acting on
$H_{\QB^{0}}(\cA_{red})$ as the screening operator
which defines the resolution of Felder type.
Due to the quotient by $Im\QB^{0}$, we have the following
identifications\footnote{This type of bosonization has been used in
ref. \cite{Sa}, in which the resolution of the Fock modules are
discussed. };
\eqn
(k+1)\partial\tc_{1}~&=&(\tH_{1}\tc_{1}), \CR
(k+1)\partial\tG_{2}~&=&(\tH_{2}\tG_{2}),
\enn
in $H_{\QB^{0}}(\cA_{red})$. Hence the ghost fields $\tc_{1}$
and $\tG_{2}$ are realized as the vertex operators
\eqn
\tc_{1}~&=& \exp \biggl(~\I\sqrt{2\over{k+1}}\phi_1
                 \biggr), \CR
\tG_{2}~&=& \exp \biggl( {\I\over\sqrt{2(k+1)}}(\phi_1
                 -i\phi_2 )\biggr),
\enn
Thus we have a free field realization of $H_{\QB^{0}}(\cA_{red})$
in terms of $\phi_1,~\phi_2,~\psi$ and $\dpsi$.
The desired BRST cohomology should be taking the $\QB^{1}$
cohomology on this space, which takes the form
\eq
\QB^{1} = - \exp \biggl(~\I\sqrt{{2\over{k+1}}} \phi_1
\biggr) - \dpsi \exp \biggl( {\I\over\sqrt{2(k+1)}}
(\phi_1 -i \phi_2 )\biggr).
\en
The first term may be interpreted as the standard screening
operator on the \lq\lq matter" sector which defines the Virasoro
minimal model. On the other hand, the second term introduces
a new feature in the free field approach to the minimal
model coupled to the two dimensional gravity.
It is very important to understand the role of the
second screening operator in the algebraic structure of
non-critical string theory.

It is possible to generalize our method to affine Lie algebras
$sl(n+1|n)^{(1)}$, which has been suggested to give the non-critical
$W_{n}$-strings.
This subject will be discussed in a forthcoming paper \cite{ItKa}.

The authors would like to thank S.-K. Yang for useful discussions.
Most part of the work of H.K. was done during his stay at
DAMTP, Univ. of Cambridge. He would like to thank
Prof. Peter Goddard for his hospitality and
the Nishina Memorial Foundation for financial support.
The work of K.I. is partially supported by University of Tsukuba
Research Projects.

\end{document}